\documentclass[12pt]{article}

\usepackage{axodraw,cite}

\newcommand{\dimplus}[2]{$(#1\!+\!#2)$-dimensional}
\newlength{\extraspace}
\newlength{\extraspaces}
\newcommand{\be}{\begin{equation}
\addtolength{\abovedisplayskip}{\extraspaces}
\addtolength{\belowdisplayskip}{\extraspaces}
\addtolength{\abovedisplayshortskip}{\extraspace}
\addtolength{\belowdisplayshortskip}{\extraspace}}
\newcommand{\ee}{\end{equation}}

\newcommand{\newsection}[1]{
\vspace{15mm}
\pagebreak[3]
\addtocounter{section}{1}
\setcounter{subsection}{0}
\setcounter{footnote}{0}
\noindent
{\Large\bf \thesection. #1}
\nopagebreak
\medskip
\nopagebreak}

\newcommand{\ba}{\begin{eqnarray}
\addtolength{\abovedisplayskip}{\extraspaces}
\addtolength{\belowdisplayskip}{\extraspaces}
\addtolength{\abovedisplayshortskip}{\extraspace}
\addtolength{\belowdisplayshortskip}{\extraspace}}

\newcommand{\reff}[1]{(\ref{#1})}
\renewcommand{\d}{\partial}

\begin{document}
\begin{titlepage}
\begin{center}

{\hbox to \hsize{\hfill PUPT-1975}}
{\hbox to \hsize{\hfill SPIN-2001/04}}

\bigskip

\vspace{6\baselineskip}

{\Large \bf CFT and Entropy on the Brane}

\bigskip

\bigskip
\bigskip

{\large \sc Ivo Savonije\footnote{savonije@phys.uu.nl}\\[1mm] {\it and}\\[2mm]
Erik Verlinde\footnote{erikv@feynman.princeton.edu}}\\[1cm]

{\it $^1$ Spinoza Institute, University of Utrecht, Utrecht, The
Netherlands}\\[3mm]

{\it $^2$ Physics Department, Princeton University, Princeton, NJ 08544}\\[3mm]

\vspace*{1.5cm}
\large{

{\bf Abstract}\\
}

\end{center}
\noindent
We consider a brane-universe in the background of an
Anti-de Sitter/Schwarschild geometry. We show that the induced
geometry of the brane is exactly given by that of a standard radiation
dominated FRW-universe. The radiation is represented by a strongly
coupled CFT with an $AdS$-dual description. We show that when the brane
crosses the horizon of the $AdS$-black hole the entropy and temperature
are simply expressed in the Hubble constant and its time
derivative. We present formulas for the entropy of the CFT which are
generally valid, and which at the horizon coincide with the FRW
equations.  These results shed new light on recently proposed entropy
bounds in the context of cosmology.

\end{titlepage}
\newpage

\newsection{Introduction}

Recently the holographic principle was studied in a
Friedmann-Robertson-Walker (FRW) universe filled with a conformal
field theory (CFT) with a dual anti-de~Sitter ($AdS$)
description~\cite{erik}, see
also~\cite{larsen,lin,nojiri,wang,klemm,brustein}. An interesting and
surprising relationship was found between the FRW equations
controlling the cosmological expansion and the formulas that relate
the energy and entropy of the CFT. The aim of the present paper will
be to shed further light on this coincidence by studying the
CFT/FRW-cosmology from a Randall-Sundrum type brane-world
perspective~\cite{rs,brane}.

Brane cosmology has previously been studied from an $AdS$/CFT perspective in
\cite{gubser,ads}. Following these papers we describe the CFT dominated
universe as a co-dimension one brane, with fixed tension, in the
background of an $AdS$-black hole. In this description the movement of
the brane turns out to be exactly described by the standard Friedmann
equation in which the size of the universe directly corresponds to the
distance of the brane to the center of the black hole. The brane
starts out inside the black hole, it passes through the horizon and
keeps expanding until it reaches a maximal radius, after which it re-contracts
and falls back into the black hole. From the $AdS$-perspective there are two
special moments, one in the early and one in the late universe,
when the brane crosses the horizon. The main goal of this paper is to show
that at those moments the entropy density on the brane takes a special value
given in terms of the Hubble constant and Newtons constant. Furthermore, at
these times the Friedmann equation turns into an equation that expresses the
entropy density in terms of the energy density and exactly coincides
with a generalized form of the Cardy formula for the entropy of the CFT.

We begin by presenting the brane description of a CFT-dominated cosmology in
section~2. The dimension $d=n\!+\!1$ of the brane-universe
will be taken to be arbitrary, but its relation with the dimension
$D=d\!+\!1$ of the $AdS$ space is of course fixed. In section~3 we argue
that the radiation on the brane can be identified with the  CFT dual
to the $AdS$-space and use this fact to fix the normalization of
Newtons constant and derive the FRW equations. The entropy density and
temperature of the CFT at the moment that the brane crosses the
horizon are calculated in section~4. We find that these quantities
have a simple expression in terms of the Hubble constant and its
time-derivative. In section~5 we derive the entropy formulas for the
CFT and show the correspondence with the FRW equations. Finally,
sections~6 and~7 contain some concluding remarks.


\newsection{Brane cosmology}
\label{sec:brane}

We consider an \dimplus{n}{1} brane with a constant tension
in the background of an \dimplus{n}{2} $AdS$-Schwarzschild black hole.
Following the $AdS$/CFT prescription \cite{klebanov,wit} we regard the
brane as the boundary of the $AdS$-geometry. An important difference
is, however, that now the location and the metric on the boundary are,
at least partly, dynamical. The movement of the brane is described by
the boundary action
\be
  {\cal L}_{b} = \frac{1}{8\pi\mathbf{G}_N}\int_{\partial{\mathcal
  M}} \sqrt{g} \; \mathcal{K} + \frac{\kappa}{8\pi\mathbf{G}_N}
  \int_{\partial{\mathcal M}} \sqrt{g} \, .
\ee
Here $\mathcal{K}\equiv\mathcal{K}_i^{\;i}$ is the trace of the extrinsic
curvature, $\kappa$ is a parameter related to the tension of the
brane, $\mathbf{G}_N$ is the \dimplus{n}{2} bulk Newton constant, $g$
is the determinant of the induced metric and $\partial {\mathcal{M}}$
denotes the surface of the brane.
The equation of motion of the brane that follows from this Lagrangian is
\be
  \label{eq:extrinsic}
 \mathcal{K}_{ij} = {\kappa\over n} \, g_{ij}^{\mbox{\tiny induced}}.
\ee
This equation implies that $\partial{\mathcal M}$ is a surface of
constant extrinsic curvature.

The bulk action is given by the \dimplus{n}{2} Einstein action with
cosmological term. The $AdS$-Schwarzschild metric provides a solution
of the bulk equations of motion and can be written in the following
form,
\begin{eqnarray}
  \label{eq:metric}
    ds_{n+2}^2 &\! =\! & \frac{1}{h(a)} da^2 - h(a) dt^2 + a^2
    d\Omega^2_{n} \, , \\
    h(a) &\! =\! & \frac{a^2}{L^2} + 1 - \frac{\omega_{n+1}M}{a^{n-1}},
\end{eqnarray}
where
\be
  \omega_{n+1} = \frac{16\pi \mathbf{G}_N}{n\mbox{Vol}(\mathbf{S}^n)}.
\ee
In these equations, $L$ is the curvature radius of $AdS$. The
pre-factor $\omega_{n+1}$ is chosen such that $M$ is the mass of the
black hole as measured by an observer who uses $t$ as his time coordinate.

Our aim is to find the spherically symmetric solutions corresponding to
a homogeneous and isotropic induced metric on the brane.
Let us parameterize the location of the brane by giving $a$ as a function
of the $AdS$-time $t$.
Equivalently, we may introduce a new time parameter $\tau$ and
specify the functions
\be
  a = a(\tau), \  t=t(\tau) .
\ee
We will choose the time parameter $\tau$ such that the following relation
is satisfied,
\be
  \label{eq:slice}
  \frac{1}{h(a)} \left(\frac{da}{d\tau}\right)^2 - h(a) \left(
  \frac{dt}{d\tau} \right)^2 = -1 .
\ee
This condition ensures that the induced metric on the brane takes the
standard Robertson-Walker form,
\be
  \label{eq:rw}
  ds_{n+1}^2 = -d\tau^2 + a^2(\tau) d\Omega^2_{n} \, .
\ee
We note that the size of the \dimplus{n}{1} universe is determined by the
radial distance, $a$, from the center of the black hole.

The extrinsic curvature, $\mathcal{K}_{ij}$, of the brane can be
straightforwardly calculated and expressed in term of the functions
$a(\tau)$ and $t(\tau)$. One then finds that the equation of
motion~\reff{eq:extrinsic} translates into
\be
  \label{eq:constraint}
  {dt\over d\tau} = \frac{\kappa a}{h(a)}.
\ee
In the following we will tune the \dimplus{n}{1} cosmological constant
to zero by setting $\kappa = 1/L$. Combining \reff{eq:constraint} with
\reff{eq:slice} leads to an equation that looks suspiciously like the
Friedmann equation for a radiation dominated universe,
\be
\label{eq:fm}
H^2 = - \frac{1}{a^2} + \frac{\omega_{n+1}M}{a^{n+1}} .
\ee
In this equation, $H\equiv\dot{a}/a$ is the Hubble `constant' and the
dot denotes differentiation with respect to the cosmological time $\tau$.
For future purpose, we also give the equation for the time derivative of $H$,
\be
\label{eq:fm2}
\dot{H} = \frac{1}{a^2} - \frac{(n+1)}{2} \frac{\omega_{n+1}M}{a^{n+1}},
\ee
which is simply obtained by differentiating \reff{eq:fm}.

\begin{figure}[tb]
  \begin{center}
   \begin{picture}(0,150)
     \Photon(-70,0)(70,0){-2}{15.5}
     \Photon(-70,140)(70,140){2}{15.5}
     \Line(-70,0)(-70,140)
     \Line(70,0)(70,140)
     \Line(-70,0)(70,140)
     \Line(70,0)(-70,140)
     \CArc(-55,70)(100,-43,43)
     \LongArrow(85,70)(48,70)
     \Text(87,72)[l]{Brane worldline}
     \LongArrowArcn(85,170)(55,270,215)
     \Text(87,115)[l]{Future singularity}
     \LongArrowArc(85,-30)(55,90,145)
     \Text(87,25)[l]{Past singularity}
     \LongArrowArcn(-85,-1)(70,90,38)
     \LongArrowArc(-85,141)(70,270,322)
     \Text(-87,72)[r]{Horizon}
     \GCirc(37.5,32.5){2}{0}
     \GCirc(37.5,107.5){2}{0}
   \end{picture}
\caption{\small Penrose diagram of an $AdS_{n+2}$-Schwarzschild black hole
with the trajectory of the brane. The brane originates in the past
singularity, expands to a certain size and subsequently falls
into the future singularity as it re-collapses. The dots indicate
the moments when the brane crosses the black hole horizon.}
    \label{fig:penrose}
  \end{center}
\end{figure}
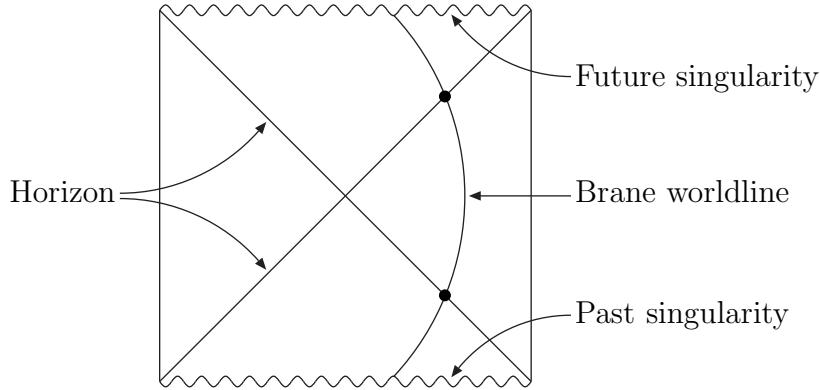


\newsection{CFT on the brane}
\label{sec:adscft}

We now want to identify the equation of motion \reff{eq:fm} with the
\dimplus{n}{1} Friedmann equation. In particular, we will argue that
the radiation can be identified with the finite temperature CFT that is dual
to the $AdS$-geometry. To do so, we interpret the last term on the r.h.s.
as the contribution of the energy density $\rho$ of the CFT times
the \dimplus{n}{1} Newton constant $G_N$.
In the brane-world scenario the relation between the Newton constant
$\mathbf{G}_N$ in the bulk and the Newton constant $G_N$ on the brane
is given by
\be
\label{GG}
\mathbf{G}_N=\frac{G_N L}{(n-1)}.
\ee
One possible way to derive this fact is to add a small amount of stress energy
on the brane and determine how it effects the equation of motion. This same
relation is, as we will discuss, also consistent with the identification of
the radiation with the dual CFT.

In \cite{witten} it was argued that the energy, entropy and temperature
of a CFT at high temperatures can be identified with the mass, entropy and
Hawking temperature of the $AdS$ black hole \cite{hawking}. The CFT lives on
a space-time which, after Euclidean continuation, has the topology of
$\mathbf{S}^1\times \mathbf{S}^n$ and whose geometry is identified with
the asymptotic boundary of the Euclidean $AdS$-black hole.
We remind the reader that the standard GKPW
prescription~\cite{klebanov,wit} of the $AdS$/CFT
correspondence~\cite{malda} only
fixes the conformal class of the CFT metric. It thus specifies only the
ratio of the radius of the $n$-sphere to the Hawking temperature but
does not fix the overall scale of the boundary metric.
One is therefore free to re-scale the metric as one wishes.
It is important to note, however, that such a
rescaling does also affect the energy and temperature of the CFT.

To make this more precise, let us consider the asymptotic form
of the $AdS$-Schwarschild metric. We have
\be
\lim_{a\to \infty} \left[\frac{L^2}{a^2}\, ds^2_{n+2}\right]
= -dt^2 + L^2 d\Omega_n^2 \, ,
\ee
from which we see that the CFT time is equal to the $AdS$
time $t$ only when the radius of the  spatial sphere is set equal to $L$.
Therefore, if we want the sphere to have a radius equal to say $a$, the
CFT time will be equal to $at/L$. The same factor $a/L$ then appears in the
relation between the energy $E$ and the black hole mass $M$. One thus
finds that the energy for a CFT on a sphere with radius $a$, of volume
$$
V=a^n{\rm Vol}(\mathbf{S}^n),
$$
is given by
\be
\label{eq:merel}
E=M{L\over a} .
\ee
Note that the total energy $E$ is not constant during the cosmological
expansion, but decreases like $a^{-1}$. This is consistent with the
fact that for a CFT the energy density,
$$
\rho={E\over V},
$$
scales like $a^{-(n+1)}$.
Inserting the relation \reff{eq:merel} combined with \reff{GG} into
the equation of motion \reff{eq:fm} leads to
\be
\label{eq:friedmann}
H^2 = - \frac{1}{a^2} + \frac{16\pi G_N}{n(n-1)}\rho \, .
\ee
This is the standard Friedmann equation with the appropriate normalization
for both terms. By differentiating once with respect to $\tau$ and using
the fact that $\dot{\rho}=nH(\rho+p)$, one derives the second FRW equation,
\be
\label{eq:friedmann2}
\dot{H} = \frac{1}{a^2} - \frac{8\pi G_N}{(n-1)} \left( \rho + p
\right) ,
\ee
which is equivalent to \reff{eq:fm2}.
An observer on the brane, who knows nothing about the $AdS$-bulk gravity,
just notices the normal cosmological expansion.
The brane description contains more information, since it also knows
about the size of the $AdS$-black hole.

The movement of the brane in the $AdS$-black hole background is depicted in
the Penrose diagram in figure \ref{fig:penrose}.
The diagram represents the full
geodesically complete black hole geometry including the asymptotic region
$a\to \infty$. If one wants to take the brane as the
real boundary, one has to cut away the part to the right of the brane.
We see that the brane indeed starts inside the black hole at
the past singularity and then, as it expands, it moves away from $a=0$.
At late times it does the opposite. The points  where the brane
crosses the black hole horizon  will play a central role in the
following discussion and have been marked in the figure. These moments
are clearly distinguished from the $AdS$-perspective, even though
nothing special happens to the induced geometry on the brane. So what
do these moments mean for an observer on the brane?


\newsection{Entropy and temperature at the horizon}
\label{horizon}

Let us now consider the points at which the brane crosses the
horizon. The horizon of the $AdS$-black hole
is located at radius $a=a_H$, where $a_H$ is the largest
solution to the equation $h(a)=0$, i.e.
\be
\frac{a_H^2}{L^2} + 1 - \frac{\omega_{n+1}M}{a_H^{n-1}}=0 \, .
\ee
>From this equation and the equation of motion \reff{eq:fm}, one immediately
concludes that the Hubble constant at the horizon obeys
$$
H^2={1\over L^2},
$$
and hence $H=\pm{1/L}$ depending on whether the brane is expanding
or contracting.

Next, let us consider the entropy density. According to \cite{witten},
the entropy of the CFT is equal to the Bekenstein-Hawking entropy of the
$AdS$-black hole, which is given by the area of the horizon measured in
bulk planckian units.
The total entropy may thus be expressed as
\be
S={V_H \over 4\mathbf{G}_N},
\ee
where $V_H$ is the area of the horizon,
$$
V_H \equiv a_H^n {\rm Vol}(\mathbf{S}^n).
$$
Note that the area of an $n$-sphere in $AdS$ equals the volume of the
corresponding spatial section for an observer on the brane.
The total entropy $S$ is constant during the cosmological
evolution but the entropy density,
$$
s={S\over V},
$$
of course varies with time. It equals
\be
\label{sofa}
s= (n-1) {a^n_H\over 4G_N L a^n},
\ee
where we made use of the relation \reff{GG}.
What makes the moments that the brane crosses the horizon
special is that the entropy density is given by a simple multiple of
the Hubble constant $H$.  At the horizon $V=V_H$ and hence the entropy density on the brane is $s=1/4\mathbf{G}_N$. Now, using the relation \reff{GG}
and the fact that $H=1/L$ one finds that the entropy density equals
\be
\label{s}
s=(n-1){H\over 4G_N}, \qquad\quad\mbox{at $a=a_H$} \, .
\ee
The significance of this relation will be further discussed below.

Also the temperature turns out to have a special value at the horizon.
The Hawking temperature measured by an observer who uses
$t$ as his time coordinate is \cite{witten,gubser}
\be
T_H={h'(a_H)\over 4\pi},
\ee
where the prime denotes differentiation with respect to $a$.
Since the CFT time differs from $t$ by a factor $a/L$ the CFT-temperature $T$
will differ from the Hawking temperature
$T_H$ by the same $a$-dependent factor,
\be
T=T_H {L\over a}.
\ee
Using the explicit form of $h'(a_H)$ and using the fact that $h(a_H)=0$,
we eventually find
\be
\label{tofa}
T= \frac{1}{4\pi a}\left((n+1)\frac{a_H}{L}+(n-1)\frac{L}{a_H}\right).
\ee
Now, from the derivation of the brane equation of motion, it follows that
the quantities $H^2$ and $-h(a)/a^2$ only differ by a constant and
therefore, at the horizon where $h(a_H)=0$, we have that
$\dot{H}=- h'(a_H)/2a_H$. This can be used to show that the
temperature at the horizon may be expressed in the Hubble constant
$H$ and its time derivative $\dot{H}$ as
\be
\label{T}
T=- {\dot{H}\over 2\pi H}, \qquad\quad\mbox{at $a=a_H$} \, .
\ee


\newsection{Entropy formulas and FRW equations}
\label{entropy}

The above relations between the entropy density and temperature on the one
hand, and the Hubble constant, its time derivative and Newtons constant on
the other are valid only when the brane crosses the horizon.
However, since the entropy density, temperature and energy density all vary in
a precisely prescribed manner as a function of the radius $a$, these
relations imply a set of entropy formulas that remain valid at all times.

Before making this point clear, let us first briefly discuss some
basic thermodynamics. The first law of thermodynamics,
$$
TdS=dE+pdV ,
$$
can after some straightforward manipulations
be rewritten in terms of the entropy and energy densities $s$ and $\rho$ as
\be
\label{dsdro}
Tds=d\rho+n(\rho+p-Ts){da\over a},
\ee
where we used $dV=nVda/a$.
The combination $(\rho+p-Ts)$ is in most standard textbooks on
cosmology \cite{weinberg,kolb}
assumed to vanish, which is equivalent to saying that the
entropy and energy are purely extensive. But let us now compute it for
the CFT.
The energy density is given by
\be
\rho ={ML\over a^{n+1}{\rm{Vol}(\mathbf{S}^n)}}.
\ee
For our purpose, it is convenient to rewrite $\rho$ in terms of
the horizon radius $a_H$ using $h(a_H)=0$. This gives
\be
\rho ={n a_H^n \over 16\pi \mathbf{G}_N a^{n+1}}
\left({L\over a_H}+{a_H\over L}\right).
\ee
The pressure follows from $\rho$ through the equation of state  $p=\rho/n$.
Combined with \reff{sofa} and \reff{tofa}, one gets
\be
\label{defgamma}
{n\over 2}(\rho+p-Ts)={\gamma\over a^2} \, ,
\ee
where the quantity $\gamma$ is given by
\be
\label{gamma1}
\gamma=\frac{ n (n-1) a_H^{n-1}}{16 \pi G_N a^{n-1}} \, .
\ee
Equation \reff{defgamma} may be regarded as the definition of $\gamma$.
Physically one can think of $\gamma$ as describing the response of
the energy density under variations of the radius $a$ or, more precisely,
the spatial curvature $1/a^2$. It thus represents the geometrical
Casimir part of the energy density.

We are now ready to present the main entropy formula for CFT's with an
$AdS$ dual. In \cite{erik} an entropy formula was already derived and
expressed in terms of the total energy and entropy.
Here we will give the local version in terms of densities. From the given expressions for the entropy
density $s$, energy density $\rho$ and $\gamma$, one finds that $s$ may be expressed as
\be
\label{cardy}
s^2=\left(4\pi\over n\right)^2\gamma \left(\rho-{\gamma\over a^2}\right).
\ee
As noted in \cite{erik}, this formula resembles the Cardy formula of
a \dimplus{1}{1} CFT but is valid for all spatial dimensions $n$.

The formulas \reff{defgamma} and \reff{cardy} are valid at all times.
It will be interesting, however, to study these formulas at the special
time when the brane crosses the horizon. First we note that at
that time the Casimir quantity $\gamma$ equals
\be
\label{gamma}
\gamma={n(n-1)\over 16 \pi G_N}, \qquad\mbox{at $a=a_H$} \, .
\ee
Let us now consider the entropy formula \reff{cardy}.
By making the identifications \reff{s} and \reff{gamma}
one sees that this formula exactly reproduces the Friedmann equation!
Similarly, one finds that equation \reff{defgamma} reduces to the second
FRW equation for $\dot{H}$ by making the same substitutions
for $s$ and $\gamma$ and replacing the temperature $T$ by the
r.h.s. of \reff{T}. In fact, the equations \reff{defgamma} and \reff{cardy}
are equations of state of the CFT and in principle have an
interpretation that is independent of gravity or cosmology. It seems
therefore rather surprising that the Friedmann equation knows about the
thermodynamic properties of the CFT.


\newsection{Euclidean brane cosmology}
\label{euclides}

In principle one can use the present setup to calculate the
correlation functions of operators in the CFT/FRW cosmology,
in particular the stress energy tensor, using the same methods as in
the standard $AdS$/CFT setup. This would for example give information
about fluctuations in the energy density in the early universe.
As described above, the brane starts out as a point in the past singularity of
the black hole.
The presence of this singularity may lead to problems in performing
these calculations in Minkowski signature.
On the gravity side a singularity is associated with the UV properties of the
theory, i.e. to very high energies. However, through the UV/IR-connection
\cite{susskind} known from $AdS$/CFT, on the field theory side this in fact
corresponds to the IR, i.e. to very low energies. As it is the CFT that
describes the matter in the universe, this seems strange since
conventionally one associates the UV with the early universe.

\begin{figure}[tb]
  \begin{center}
   \begin{picture}(0,100)
     \BCirc(0,50){40}
     \GCirc(0,50){1}{0}
     \CArc(75,50)(55,149,211)
     \LongArrow(65,65)(28,65)
     \Text(67,67)[l]{Brane worldline}
   \end{picture}
    \caption{\small Diagram of Euclidean $AdS_{n+2}$-Schwarzschild with the
     trajectory of the brane. The horizon is represented by the dot in the
     middle of the diagram; only the region $a\geq a_H$ is drawn. The
     brane originates at spatial infinity, collapses to a certain
     miminal size and subsequently re-expands. It remains outside of the
     black hole horizon during the entire evolution.}
    \label{fig:Euclid}
  \end{center}
\end{figure}
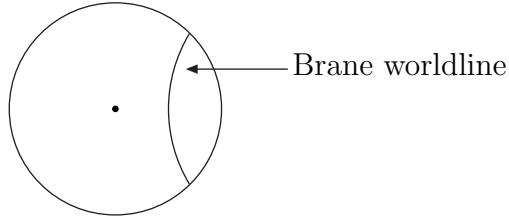

To calculate correlation functions one can circumvent this problem by
analytically continuing to the Euclidean setup.
So let us briefly discuss how to describe the Euclidean FRW universe as
a brane in an Euclidean $AdS$-Schwarzschild background.
Going through the calculation in a similiar way as performed above, one
arrives at the following Friedmann equation
\be
  H^2_{\rm E} = \frac{1}{a^2} - \frac{16\pi G_N}{n(n-1)} \rho.
\ee
>From this one easily deduces that the universe, when regarded in
Euclidean time, undergoes a reverse evolution, starting out very big,
collapsing to a minimal size and subsequently re-expanding. This is
depicted in figure \ref{fig:Euclid}. From the CFT point of view, this
means that the universe starts in the far UV, then cools down to a
certain minimum temperature after which it re-heats. Note that in
this case, the brane does not cross the horizon at all.

\newsection{Conclusion}
\label{sec:conclusion}

In \cite{erik} it was argued that the discovered relation between
the FRW equations and the entropy formulas sheds light
on the meaning of the holographic principle in a cosmological setting
\cite{hubble}.
Indeed, it was suggested that the values for $s$ and $T$ on the horizon
should be regarded as bounds on these respective quantities. Although we
still have no proof of this fact, we would like to present some further
arguments in favor of this. At the moment when the brane crosses the
horizon, the quantity $\gamma$ is essentially equal to the inverse
Newton constant.
This means that the response of the energy density to a variation
of the curvature is comparable to that of the Einstein action itself.
Namely, from \reff{defgamma} and \reff{gamma} one finds
\be
a\left({\d \rho\over\d a}\right)_{\! s} = {-n(n-1)\over 8\pi G_N a^2}
\, , \qquad\quad \mbox{at $a=a_H$}.
\ee
The right hand side also gives the contribution of the spatial curvature in
the equation of motion.
Clearly, when this is the case one should reconsider the validity of the
usual formulation of gravity, since quantum effects
(the Casimir energy density) are of the same order as the spatial curvature.
This suggests that a classical description of the geometry of the universe
may no longer be well defined and one has to go over to a different, more
fundamental formulation of the theory.  We have indeed noticed that, at the
transition points, the laws that govern the gravitational evolution and
the entropy and energy expressions for the CFT, that describes the radiation,
merge in a surprising way. This indicates that  both sets of equations have
a common origin in a single underlying fundamental theory.
\bigskip

\begin{flushleft}
{\sc Acknowledgments}
\end{flushleft}

\noindent
I.S. wishes to thank A.~Boyarsky, Y.-H.~He, M.~Parikh and M.~Spradlin
for helpful discussions. The research of I.S. is financially supported
by the Dutch Science Foundation (NWO).


\end{document}